\documentclass[manuscript]{aastex}

\shorttitle{High-collimated Water Maser Bipolar Outflow}
\shortauthors{Chibueze, J. O et al.}

\begin{document}


\title{A Highly collimated Water Maser Bipolar Outflow in the Cepheus A HW3d Massive Young Stellar Object }


\author{James O. Chibueze\altaffilmark{1,2}, Hiroshi Imai\altaffilmark{1}, Daniel Tafoya\altaffilmark{1}, Toshihiro Omodaka\altaffilmark{1}, Osamu Kameya\altaffilmark{3}, Tomoya Hirota\altaffilmark{3}, Sze$-$Ning Chong\altaffilmark{1}, Jos\'e M. Torrelles\altaffilmark{4}}


\altaffiltext{1}{Department of Physics and Astronomy, Graduate School of Science and Engineering,\\
	 Kagoshima University, 1-21-35 Korimoto, Kagoshima 890-0065, Japan.}
\altaffiltext{2}{Department of Physics and Astronomy, Faculty of Physical Sciences,\\
	 University of Nigeria, Carver Building, 1 University Road, Nsukka, Nigeria.}
\altaffiltext{3}{Mizusawa VLBI Observatory, National Astronomical Observatory of Japan, 2-21-1 Osawa, Mitaka, Tokyo 181-8588, Japan.}
\altaffiltext{4}{Instituto de Ciencias del Espacio (CSIC)-UB/IEEC, Facultat de F\'{\i}sica, \\ Universitat de Barcelona,
	 Mart\'{\i} i Franqu\`{e}s 1, E-08028 Barcelona, Spain.}

\email{james@milkyway.sci.kagoshima-u.ac.jp}


\begin{abstract}
We present the results of multi-epoch very long baseline interferometry (VLBI)
water (H$_{2}$O) maser observations carried out with the VLBI Exploration of Radio Astrometry (VERA) toward the Cepheus A HW3d object. We measured for the first time relative proper motions of the H$_{2}$O maser features, whose spatio-kinematics traces a compact bipolar outflow. This outflow looks highly collimated and expanding through $\sim$ 280 AU (400 mas) at a mean velocity of $\sim$ 21~km~s$^{-1}$ ($\sim$ 6 mas~yr$^{-1}$) without taking into account the turbulent central maser cluster. The opening angle of the outflow is estimated to be $\sim$ 30$^{\circ}$. The dynamical time-scale of the outflow is estimated to be $\sim$ 100 years. Our results provide strong support that HW3d harbors an internal massive young star, and the observed outflow could be tracing a very early phase of star formation. We also have analyzed Very Large Array (VLA) archive data of 1.3 cm continuum emission obtained in 1995 and 2006 toward Cepheus A. The comparative result of the HW3d continuum emission suggests the possibility of the existence of distinct young stellar objects (YSOs) in HW3d and/or strong variability in one of their radio continuum emission components.
\end{abstract}





\section{Introduction}

The last decade has seen a lot of efforts toward understanding how massive
stars form and evolve. The
fact that massive stars have a very significant impact
on the evolution of the interstellar medium of galaxies, with strong influences
in their environments (through strong winds, expanding HII regions,
UV radiation, supernova explosions), and sometimes activating other star formation events,
underscore the importance of this research.
However, it is still among the most poorly understood topics in the
field of astronomy. This is mainly because massive stars form in highly
obscured mediums, thus, making it very difficult to observe them in their early phases. In addition, they evolve quickly (formation timescales of $\sim$ 10$^5$ yr), and form in distant clusters and associations, therefore hard to isolate
single high-mass stars for their study (see reviews by e.g., Hoare \& Franco 2007; McKee \& Ostriker 2007; Zinnecker \& Yorke 2007).
Based mostly on theoretical simulations, different attempts have
been made in proposing the formation scenarios of massive stars.
There are mainly three proposed scenarios; massive star formation
through the merging of less massive stars
(Bonnell, Bate, \& Zinnecker 1998), competitive accretion in a protocluster environment
(Bonnell \& Bate 2006), and gravitational collapse involving high-rate, disk-assisted accretion into the core which helps
to overcome radiation pressure (Yorke \& Sonnhalter 2002;
McKee \& Tan 2003; Krumholz et al. 2005, 2009).

With high brightness temperatures exceeding 10$^{10}$ K and compact nature, H$_{2}$O masers have proven to be very useful in astrophysical studies using very long baseline interferometry (VLBI) with
milliarcsecond (mas) angular resolution, particularly in identifying present sites of high-mass star formation in molecular clouds for studying the very vicinity of massive
young stellar object (YSO) candidates (e.g.,  Genzel et al. 1981; Torrelles et al. 2001a, 2003; Imai et al. 2002; Goddi et al. 2005, 2006; Moscadelli et al. 2006;Vlemmings et al. 2006; Surcis et al. 2011).

In this paper, we report the results of nine epochs of H$_{2}$O maser observations toward the Cepheus A (Cep~A) high-mass star-forming region using the VLBI Exploration of Radio Astrometry (VERA). We adopt the distance to Cep~A to be $\sim$ 700 pc (Johnson 1957, Moscadelli et al. 2009, Dzib et al. 2011). We focus mainly on HW3d in Cep~A, one of the 16 radio continuum objects discovered by Hughes \& Wouterloot (1984), and located $\sim$ 3$''$ south from the brightest radio continuum source in the region, HW2. HW2 clearly harbors a massive YSO in it, which is indicated by the observations of a jet and a disk,
intense magnetic fields, and powerful H$_{2}$O masers (Rodr\'{\i}guez et al. 1994; Patel et al. 2005; Curiel et al. 2006;
Jim\'enez-Serra et al. 2007; Torrelles et al. 2007, 2011;
Vlemmings et al. 2010). On the other hand, the observational properties are not so clear in the case of the other
HW objects. In fact, Garay et al. (1996), through multifrequency Very Large Array (VLA) radio continuum observations, argued that some of the HW objects are internally excited by
a YSO, while others are externally shock-excited at the interface between winds of other YSOs and molecular clumps in the region. HW3d is one of the objects that was proposed to be excited
internally by its own YSO, on the basis that the radio continuum emission
presents an elongated structure with positive spectral index, which is suggestive of a thermal jet nature. Hughes, Cohen, \& Garrington (1995) also supported it on the basis of its association with
strong hydroxyl (OH) and H$_{2}$O maser emission (Cohen et al. 1984), although they did not
find evidence for outflow activity in this object. The evidence for outflow activity is now presented in this paper through our VERA H$_{2}$O maser observations, giving a conclusive support that the HW3d object is internally excited by a massive YSO.

\section{Observations}


\subsection{VERA Observations}

The observations of the Cep~A H$_{2}$O masers at 22.235080 GHz with VERA were carried out in 9 epochs from 2006 May to 2007 August. Table 1 gives a summary of these observations. At each epoch, the total observation time was about 8 hours, including the scans on the calibrators (J2005+7752, BL Lac, J2015+3710). Using the advantage of the VERA's dual-beam system, Cep~A and J2302+6405 (a position reference source spatially separated by 2.19$^{\circ}$ from Cep~A) were simultaneously observed with the aim of determining the annual parallax of Cep~A. The result of the measurement of the parallax distance would be published in a separate paper.
The received signals were digitized in four quantization levels, and then divided into 16 base-band channels (BBCs) in a digital filter unit, each of which had a bandwidth of 16 MHz, corresponding to a velocity coverage of $216{\rm~km~s}^{-1}$ centered around 50~km~s$^{-1}$ with respect to the local standard of rest (LSR). One of the BBCs was assigned to the frequency of the H$_{2}$O maser emission in Cep~A while the other 15 BBCs were assigned to the continuum emission from J2303+6405 and other sources observed in the B-beam of the VERA system.

The data correlation was made with the Mitaka FX correlator. The accumulation period of the correlation was set to 1 second. The correlation outputs consisted of 512 and 64 spectral channels for the H$_{2}$O maser and reference continuum emission, respectively. A velocity spacing of 0.21~km~s$^{-1}$~ was obtained in each spectral channel for the H$_{2}$O maser emission. 

The data reduction was made with the National Radio Astronomy Observatory (NRAO) Astronomical Image Processing System (AIPS) package using standard procedures. The instrumental delay calibration was made with the scans on the calibrators. The fringe fitting and self-calibration procedures were performed for a Doppler velocity channel including a bright maser spot (velocity component) in the HW3d region of Cep~A. Their solutions were applied to all the data and then the map of the maser clusters were made. The CLEANed image cubes of the maser source were created using the beam that was synthesized from naturally weighted visibilities. The typical size of the synthesized beam was $\sim$ 1.3 mas. The H$_{2}$O maser image cubes were made with a cell size of 0.2 mas. 

Because we carried out self-calibration procedure using a bright maser spot in the HW3d region, the maser spot positions in the image cubes were measured with respect to the bright maser spot used for self-calibration mentioned above. For the wide-field mapping and objective maser spot identification, we used an automatic pipeline script which runs on the AIPS POPS environment and mainly consists of AIPS tasks and adverbs: IMAGR, IMSTAT, and SAD. The H$_{2}$O maser feature identification was done by adopting a signal-to-noise ratio cut-off of $\sim$ 8. From the Gaussian fitting errors, we estimated the accuracies of the relative positions of the maser spots to be 0.01 -- 0.20 mas in Right Ascension and 0.02 -- 0.30 mas in declination. The individual maser features were defined as clusters of maser spots or velocity components and each feature position was defined from the brightness peak of the feature (for identification method, see Imai et al. 2002).\\

We adopted special procedures to measure the absolute coordinates of the detected H$_{2}$O masers. This enables a comparative study with the 1.3~cm continuum emission map from VLA observations described in \S 2.2. The position-reference source J2303+6405 observed concurrently in the B-beam of the VERA system was not detected by applying the normal fringe fitting. To detect the weak emission of the position-reference source and thus measure the coordinates of the maser features with respect to it, we applied the inverse phase-referencing technique. This procedure involves common calibrating the group delay residuals of the A- and B-beam data using the data of the bright continuum calibrators (BL Lac, J2005+7752) observed in the B-beam. Then we did fringe fitting and self-calibration as described above using a bright maser velocity component. Subsequently, we applied all the phase calibration solutions obtained to the position-reference source (J2303+6405) data. We successfully carried out this procedure in the observation epoch of February 18, 2007, using a bright maser spot at V$_{\rm LSR} = -$6.67~km~s$^{-1}$ near HW2 and detected the position-reference source (30 mJy\slash beam) at a position offset of 15.4 mas in R.A. and 606.5 mas in declination from the J2303+6405 map center. This inversely indicates the position offset of the phase-reference maser spot from the delay-tracking center. We determined the absolute coordinates of the position-reference maser spot (maser spot used for the AIPS self-calibration procedure) to be RA(J2000) = 22$^{h}$56$^{m}$17.97745$^{s}$  DEC(J2000) = +62$^{\circ}$01\arcmin49.3784\arcsec by computing the negative of the offset of the position-reference source (J2303+6405) from the map origin. \\

There was no maser feature consistently identified in all the epochs which could be used as a reference position, therefore we adopted the following method to trace individual maser features at as many successive epochs as possible. Firstly, from any two adjacent epochs, we calculated the mean coordinate offset of the maser features at the second epoch with respected to that at the first epoch using only the maser features that were detected at both epochs. Secondly, these mean offsets were referred with respect to the earliest epoch taken on May 13, 2006. Then, these offsets were subtracted from the coordinates originally used in the individual epochs (see Torrelles et al. 2001b). In so doing we were able to register all maser feature maps and obtain a reference frame whose spatial stability at all epochs depends on that of the maser feature distribution. The map origin of the reference frame is very close to a quasi-stationary maser feature in HW3d (Feature 15 in Table 2). We tested the stability of the reference position offset using the Feature 15 (seen at the same position within 1~mas and at the same LSR velocity; see Appendix) identified at 5 epochs. The estimated proper motion in the reference frame is $\mu_{x}$ $\sim$ 0.001~(mas yr$^{-1}$) and $\mu_{y}$ $\sim$ 0.003~(mas yr$^{-1}$), and the standard deviation of the offset coordinates of the maser feature is 0.02~mas in Right Ascension and 0.05~mas in Declination. All maser offset positions in this paper are given with respect to the derived reference position offset. 

\subsection{VLA Archive Data}

In order to compare the masers imaged from our VERA observations with the radio continuum emission of HW3d, we retrieved data from the VLA archive. We found a 1.3~cm continuum data set around the same epoch of the VERA observations under the project name AC0810, taken on 2006 February 11 in the most extended A configuration of the VLA at an observation frequency of $\sim$ 22.29 GHz. The data include the two circular polarizations with an effective bandwidth of 100 MHz. It is important to note that these continuum data do not overlap with the frequency of the H$_{2}$O masers, thus they are free of contamination from the line emission. The sources
1331$+$305 and 2230$+$697 were the flux density and phase calibrators, respectively. The calibration was carried out under standard procedures outlined in the chapter 4 of the AIPS cookbook. The assumed flux density for the amplitude calibrator was 2.59 Jy, while the estimated flux density of the phase calibrator was 0.51 Jy. The total time on source was $\sim$ 4 hours, which yielded an RMS noise in the final image of $\sim$52~$\mu$Jy beam$^{-1}$
using naturally weighted visibilities. 




\section{Results}

\subsection{Spatio-kinematics of H$_{2}$O Masers and the Morphology of the Radio Continuum Emission in HW3d}

We detected H$_{2}$O maser clusters corresponding to all the sub-regions R1-R8 around HW2, previously reported by Torrelles et al. (2011), as well as maser clusters associated with HW3d. In this paper, we will concentrate on the results obtained toward HW3d, presenting for the first time H$_{2}$O maser proper motion measurements in this object.

The VLA 2006 data turned out to be very useful because they also allowed us to explore the variability of HW3d by comparing these observations
with those of VLA 1995 reported in Torrelles et al. (1998). Figure 1 shows the VLA 1.3~cm continuum map obtained from the 2006 data, showing the HW2 radio jet, the HW3c and the HW3d objects, which are located $\sim$ 3$''$ south from HW2. In what follows, we will concentrate on the HW3d object. The main properties of the HW2 thermal radio jet and the HW3c object can be found in \citet{rod94} and \citet{gar96}.

Figure 2 shows the distribution and the relative proper motions of the H$_{2}$O maser features around HW3d superposed on the VLA 1.3 continuum map of this object. The accuracy in the absolute position of the maser feature estimated from the astrometric data analysis is better than 1~mas. On the other hand, the accuracy in the absolute position of the continuum emission observed with the VLA is estimated to be $\sim$ 50~mas.
We find that the H$_{2}$O maser features are distributed on a linear structure of $\sim$ 400 mas (280 AU) in size, well aligned with the elongation direction of the radio continuum emission of HW3d ($\sim$ 500 mas in total length). Within this linear structure, we identify three main clusters of masers, one located at the center position (hereafter C cluster), and the other two in opposite directions with respect to the center, at $\sim$ (--170 mas, +40 mas) (hereafter NW cluster) and $\sim$ (+220 mas, --80 mas) (hereafter SE cluster), respectively. The maser proper motions of these three groups of masers indicate the presence of a bipolar outflow moving outward from the central positions along the major axis of the radio continuum emission. In fact, while the masers of the NW cluster are moving toward the northwest, those of the SE cluster are moving toward the southeast. These bipolar motions are also observed in the masers of the central cluster at scales of $\sim$ 4 mas (see Figure 2 and Table 2). The spatial distribution of the masers and the mean value of the proper motions without considering the turbulent central cluster ($\sim$ 6 mas/yr or $\sim$ 21~km~s$^{-1}$) indicate that in HW3d there is a collimated bipolar outflow driven by an internal (probably) massive YSO that we propose to be located very close to the central position of cluster C. The fact that this cluster has the highest radial velocity dispersions ($\sim$ 20~km~s$^{-1}$; Table 2) of all the observed maser clusters in HW3d, also supports this C cluster as the main center of activity containing a driving YSO. The estimated dynamical time-scale of this outflow is $\sim$ 100 years. The fact that the radio continuum emission of HW3d is elongated along the direction of the outflow masing motions suggests a radio jet nature for this object, supporting the interpretation given by Garay et al. (1996) based on spectral index measurements.\\

There is in addition a fourth group of masers located at $\sim$ (+135 mas, --39 mas), but their proper motions toward the north-west (see Fig. 2 and Table 2) cannot be explained within a single bipolar outflow scenario excited by a single YSO close to the (0,0) position. The possibility of multiplicity of YSOs in this region is considered in \S 4.
We estimated the opening angle of the outflow using the deconvolved size of the radio jet of HW3d ($\sim$ 325 $\times$ 99 mas), giving a value of $\sim$ 30$^{\circ}$.

\subsection {Position and Velocity Variance/Covariance Matrix Analyses of the HW3d H$_{2}$O Maser Spatio-kinematics}

We carried out the position and velocity variance/covariance matrix analyses of the H$_{2}$O maser spatio-kinematics to test the existence of an outflow in the H$_{2}$O maser region. The positional and kinematical essentials were extracted using the position and velocity variance/covariance matrix (PVVCM) diagonalization technique (Bloemhof 1993; 2000) for the whole maser feature proper motions. We estimated the uncertainties associated with the derived eigenvectors and eigenvalues from their standard deviations calculated from the Monte Carlo simulation for the VVCM  diagonalization using velocity vectors randomly distributed around the observed values within their estimated errors (Imai et al. 2006). This technique is not a model fitting approach, having no free parameters. It is a fully objective, analytic tool, generally composed of the following elements;
\begin{equation}
\sigma_{ij} = \frac{1}{N - 1} \sum_{n=1}^{N} (v_{i,n} - \bar{v_{i}})(v_{j,n} - \bar{v_{j}}),
\end{equation}
where $i$ and $j$ denote the two dimensional space axes in the case of position, or three orthogonal space axes in the case of velocity. $n$ is the $n$-th maser feature in the collection summing up to $N$ ($= 30$). The bar indicates averaging over the maser features. The diagonalization of the position variance-covariance matrix (PVCM) gives the essentials of the maser position field, while that of the VVCM gives the essentials of the velocity field. Classification of the spherical symmetry of an outflow into spherical outflow from a cometary or a bipolar outflow can be done using the relative magnitudes of the three principal velocity variances. The direction of the major principal axis gives the outflow axis of a bipolar outflow. PVCM and VVCM provide a robust and objective means of extracting the position and kinematic essentials from maser proper motions, and also support some other spherical symmetric or asymmetric model in deriving the vital clues from masers. The position angles and inclinations are derived from the eigenvectors obtained from the diagonalized matrices. As examples, the diagonalized matrices of one PVCM and one VVCM are shown below while the results are presented in Table 3.

We applied Eq. 1 in two spatial dimensions using the position offsets of all the 30 maser features. The x- and y- axes correspond to the Right Ascension and Declination respectively.
The diagonalized PVCM in the unit of mas$^{2}$ as a $2 \times 2$ matrix is:
\begin{equation}
\left(
\begin{array}{cc}
11421.79 & -3811.39 \\
-3811.39 & 1308.13
 \end{array}
\right)
\Longrightarrow  
\left( 
\begin{array}{cc}
12697.28 & 0 \\
0 & 32.64
\end{array}
\right)
\end{equation}.

The larger eigenvalue of position dispersion is 389 times the smaller one. This indicates a very high collimation with respect to the position dispersion of the masers. The position angle, PA, of the eigenvector corresponding to the larger eigenvalue is 108$^{\circ}$.\\

To verify our interpretation, we also made the 3-dimensional VVCM diagonalization analysis of the H$_{2}$O maser proper motions, obtaining the following $3 \times 3$ diagonalized matrix in units of km$^{2}$s$^{-2}$:
\begin{equation}
\left(
\begin{array}{ccc}
85.49 & -35.62 & -7.62 \\
-35.62 & 59.99 & 5.00\\
-7.62 & 5.00 & 12.50
\end{array}
\right)
\Longrightarrow 
\left(
\begin{array}{ccc}
111.41 & 0 & 0 \\
0 & 34.90 & 0 \\
0 & 0 & 11.66 
\end{array}
\right)
\end{equation}.

The eigenvalues in the diagonalized matrix have one principal velocity variance dominating, 3 times the second largest eigenvalue. The factor of three between the largest eigenvalue with respect to the second one in the VVCM indicates collimation of water maser proper motions. The eigenvector $\nu_{\parallel}$ corresponding to the principal or largest eigenvalue corresponds to the axis of maximum internal velocity dispersion in the outflow. $\nu_{\parallel}$ lies at a position angle of 123$^{\circ}$, and it makes a small inclination angle of -5$^{\circ}$ with the sky plane (the negative sign means that the vector is rising out of the plane, pointing towards the observer). The consistency in the position angles of the largest eigenvalues (108$^{\circ}$ and 123$^{\circ}$ for PVCM and VVCM analyses, respectively) within the error estimated from VVCM (see Table 3) indicates the existence of an outflow whose major axis is aligned with the major axes of the PVCM and VVCM.

The ratio of the  second largest eigenvalue to the smallest eigenvalue characterizes the degree of transverse asymmetry in the maser outflow kinematics. The corresponding eigenvector to the second largest eigenvalue $\nu_{\perp}$ (connotes a principal axis in the perpendicular plane) gives some significant information about the outflow. The factor of three between the second largest eigenvalue and the smallest one indicates an azimuthally asymmetric collimated structure, and the cross section transverse to the main axis is an ellipsoid elongated in the direction of the eigenvector corresponding to the second largest eigenvalue. Such azimuthal asymmetry is found in the bipolar outflow in HW3d, in which $\nu_{\perp}$ lies at position angle of 33$^{\circ}$ and makes an inclination angle of 1$^{\circ}$ with the sky plane, pointing away from the observer.

\subsection{Properties of the Outflow}

In order to test the originating point of the outflow exciting the surrounding H$_{2}$O maser features, 
we performed the least-squares method for the model-fitting analysis as presented by Imai et al. (2000, 2011). This fundamentally involves the minimizing of the squared sum of the difference between the observed and model velocities, $S^{2}$.

\begin{eqnarray}
\nonumber
S^{2}& = & \frac{1}{3N_{\rm m}-N_{\rm p}}
\sum^{N_{\rm m}}_{i}
\left\{
\frac{\left[\mu_{ix}-w_{ix}/(a_{{\rm 0}}d)\right]^{2}}{\sigma^{2}_{\mu_{ix}}}+\frac{\left[\mu_{iy}-w_{iy}/(a_{{\rm 0}}d)\right]^{2}}{\sigma^{2}_{\mu_{iy}}}
+\frac{\left[u_{iz}-w_{iz}\right]^{2}}{\sigma^{2}_{u_{iz}}} 
\right\},
\label{eq:model-fit}
\end{eqnarray}

\noindent
where $N_{\rm m}$ is the number of maser features with measured proper motions, $N_{\rm p}$ the number of free parameters in the model fitting, $a_{\rm 0}=~$4.74~km~s$^{-1}$~mas$^{-1}$yr~kpc$^{-1}$ a conversion factor from a proper motion to a linear velocity, and $d$ the distance ($\sim$ 700~pc) to the maser source from the Sun, respectively. $\mu_{ix}$ and $\mu_{iy}$ are the observed proper motion components in the R.A. and declination directions, respectively, $\sigma_{\mu_{ix}}$ and $\sigma_{\mu_{iy}}$ are their uncertainties, $u_{iz}$ the observed LOS velocity, and $\sigma_{iz}$ its uncertainty. For simplicity we assume a spherically expanding outflow. The modeled velocity vector, {\boldmath{$w_{i}$}} $(w_{ix},w_{iy},w_{iz})$, is given as 

\begin{equation}
\mbox{\boldmath $w_{i}$}=
\mbox{\boldmath $V_{\rm 0}$}+V_{\rm exp}(i)\frac{\mbox{\boldmath $r_{i}$}}{r_{i}},
\end{equation}

\noindent
where {\boldmath $V_{\rm 0}$}$(v_{{\rm 0}x}, v_{{\rm 0}y}, v_{{\rm 0}z})$ is the systemic velocity 
vector of the outflow, 
\begin{eqnarray}
\nonumber
\mbox{\boldmath $r_{i}$} & = & \mbox{\boldmath $x_{i}$}-\mbox{\boldmath $x_{\rm 0}$} \\
& & \mbox{ (or $r_{ix}=x_i-x_{\rm 0}$, $r_{iy}=y_i-y_{\rm 0}$, $r_{iz}=z_i$)}, 
\label{eq:ri}
\end{eqnarray}

\begin{equation}
z_{i}=\frac{(u_{iz}-v_{{\rm 0}z})(r^{2}_{ix}+r^{2}_{iy})}
{(u_{ix}-v_{{\rm 0}x})r_{ix}+(u_{iy}-v_{{\rm 0}y})r_{iy}}, 
\label{eq:zi}
\end{equation}

\noindent
and

\noindent
\begin{equation}
u_{ix}=\mu_{ix}a_{\rm 0}d,\;\; u_{iy}=\mu_{iy}a_{\rm 0}d.
\label{eq:uix-uiy}
\end{equation}

\noindent
Eq. \ref{eq:zi} satisfies the following constraint; the obtained
position minimizes the value of $S^{2}$, 

\begin{equation}
\frac{\partial S^{2}}{\partial z_{i}}=0.
\label{eq:s2-zi}
\end{equation}

In this model, each  maser feature is radially moving from a common originating point of the outflow with an expansion velocity, $V_{\rm exp}(i)$, which is approximated by 

\begin{equation}
V_{\rm exp}(i)=\frac{(u_{ix}-v_{{\rm 0}x})r_{ix}+(u_{iy}-v_{{\rm 0}y})r_{iy}
+(u_{iz}-v_{{\rm 0}z})r_{iz}}{r_{i}}.
\label{eq:vexp}
\end{equation}

The first step of the modeling considers the three dimensional position vectors of the maser features with respect to the originating point of the outflow by assuming that the individual maser features are moving independently and radially from the originating point. The free parameters of the model fitting are the position vector of the 
originating point ($x_{0}, y_{0}$) and the systemic motion vector of the outflow on the sky 
($V_{0{\rm x}}$, $V_{0{\rm y}}$). $z_0\equiv 0$ and $V_{0z}$ = $-12$~km~s$^{-1}$ are fixed. Here we consider the possibility of a single driving source of the outflow.

Table  \ref{tab:models} gives the parameters of the best-fit model. This fitting, considering the errors (within 2 sigma), is fully consistent with the origin of the bipolar outflow being at the position of the C H$_{2}$O maser cluster supporting our interpretation given in \S 3.1 that the C cluster is the main center of activity containing a driving YSO.

Figure \ref{fig: expansion velocities} shows the distributions of the estimated expansion velocities of the maser features V$_{\rm exp}(i)$, in which the originating point of the outflow is considered to be at or very close to the position of Feature 15 (0,0) (see Imai et al. 2000, 2011). If a single expanding flow exists, the data points should be concentrated on the range of positive expansion velocities. 
The maser kinematics in Cluster C seems to be significantly contaminated by random motion. It should be noted that this analysis has large uncertainties in the estimated sign of V$_{\rm exp}$ (positive or negative) for the individual maser features close to the estimated position of the originating point because the position estimated from the model fitting using the large scale maser clusters has a large uncertainty (see Table 4). Nevertheless, we cannot discard that some of these negative V$_{\rm exp}$ are due to infalling motions. In fact, to produce these infalling motions of $\sim$ 10~km~s$^{-1}$ at distances of 0.15$''$ (105 AU), a central binding mass of $\sim$ 12~M$_{\odot}$ would be necessary, which is not an unlikely value (a B3 star has been proposed to explain the radio continuum emission of HW3d; Garay et al. 1996). Alternatively, we also cannot discard that these masers, specially those at (+135 mas, -39 mas) are excited by a close YSO other than the driving source at the center (see below, \S 4).





\section{Discussion}

Our VERA proper motion measurements of H$_{2}$O masers have shown for the
first time outflow activity in Cep~A HW3d,  implying that it harbors (at
least) a YSO driving a high-collimated  outflow at a scale of $\sim$
0.4$''$ ($\sim$ 280~AU), with a velocity of $\sim$  21~km~s$^{-1}$ (with no consideration of the proper motions in the turbulent central maser cluster) and
dynamical time-scale of $\sim$ 100 years.  The outflow has an opening angle of 30$^{\circ}$ (see \S 3.1). The internal exciting
source is probably a massive YSO (B-early star) to account for
both the observed high intensity of the H$_{2}$O masers (up to $\sim$ 100
Jy for individual maser features) and the radio continuum emission ($\sim$ 10~mJy at 1.3~cm;
e.g., Hughes \& MacLeod 1993, Hughes, Cohen, \& Garrington 1995, Garay
et al. 1996, this paper).  These characteristics are also seen in HW2, which has already been identified as a massive YSO (Patel et al. 2005; Torrelles et al. 2011), and indicate that during the process of the formation of this likely massive YSO, a
``YSO-jet" system has been formed, similar to what happens in the
formation of low-mass stars.

However, not all the observed properties of HW3d can be explained with
just  a single YSO. In fact, the VERA observations show a cluster of
H$_{2}$O masers located at $\sim$ (+135 mas, --39 mas), with proper
motion vectors that do not fit within the bipolar expanding motions
outward from the center (\S\S~3.1, 3.2, and Fig. 2). Another YSO might in principle be responsible for the excitation of the masers in that cluster. In
addition, and most importantly, by comparing the HW3d 1.3 continuum
emission observed in two epochs (1995 and 2006)  with similar angular
resolution ($\sim$ 0.1$''$), we see that this source is
variable in its total flux density($\sim$ 9 mJy [1995], $\sim$ 6.9 mJy [2006]),  and on the
other hand that the position of its peak emission has changed by
$\sim$ 0.2$''$ between these two epochs. We find that this position
shift  is highly significant, in particular when considering that the
peak emission of the other two nearby objects HW2 and HW3c  have not
changed in position in these two epochs within $\sim$ 10~mas in right ascension and declination (see Fig. 1).  In the case
that there was only a single exciting source, then the shift in the
position of the  continuum emission could be the result of the proper
motion of the YSO. However, this would correspond to a  YSO velocity of
$\sim$ 65~km~s$^{-1}$. Such a high velocity for the proper motion of the YSO itself is quite unlikely,  thus
buttressing the likelihood of the two continuum peaks representing two
different YSOs in close proximity and high flux density
variability. Alternatively, the change in the peak position could be due to internal proper motions of clumps in the jet with flux density variations. This radio continuum variability of HW3d was also
previously reported by Hughes, Cohen, \& Garrington (1995), which also invoked the
presence of different variable YSOs to explain the main
characteristics of the object.

This scenario can be tested with high-angular sensitive
EVLA, and SMA (sub)mm observations to trace the dust continuum
emission and molecular gas around the possible different YSOs associated with
HW3d, and possible proper motions of the radio jet. These observations would also be very valuable to elucidate the
nature of the YSO(s) (mass, dust and gas contents), and in
particular to detect and study the expected circumstellar disk around
the driving source of the compact bipolar outflow observed with
VERA. The fact that in Cep~A (the second closest  high-mass star
forming region after Orion) we have two very close ``YSO-jet''
systems, HW2 and HW3d (separated in the sky by $\sim$ 3$''$, $\sim$
2100 AU), gives a  unique opportunity to study their possible
different properties in terms of the mass and evolution of the
massive systems.

\section{Summary}
We have carried out VERA 9-epoch H$_{2}$O maser observations toward the radio continuum object Cep~A HW3d. We have measured the relative proper motions of 30 H$_{2}$O masers associated with HW3d, showing for the first time outflow activity in this radio continuum object. This result gives strong support that HW3d harbors (at least) an internal massive YSO. The spatio-kinematical distribution of the masers show a main bipolar structure at scales of 0.4\arcsec (280 AU) along the direction of elongation of the continuum emission and dynamical time-scale of $\sim$ 100 years, suggesting that a ``YSO-jet" system has formed in this massive object. 

The H$_{2}$O maser proper motions that do not fit within the bipolar expanding motion outward from the originating center, as well as the observed shift in the peak positions of the VLA 1.3~cm continuum emission between different epochs, may be indicating the presence of more than one exciting sources in HW3d. This possible multiplicity can be tested with SMA and EVLA high-angular resolution observations.



\acknowledgments

We would like to thank our anonymous referee for the very careful and valuable report on our manuscript. We gratefully acknowledge support of staff members of VERA project and the Astrophysics Group of Kagoshima University. Our heartfelt appreciation goes to the students who have played some role in the VERA array operation at the four stations. J.O.C. has been financially supported by the foreign scholarship of the Ministry of Education, Culture, Sports, Science, and Technology of Japan. JMT acknowledges the support from MICINN (Spain) AYA2011-30228-C03 and AGAUR (Catalonia) 2009SGR1172 grants.

\clearpage



\begin{figure}
\begin{center}
\epsscale{0.8}
\plotone{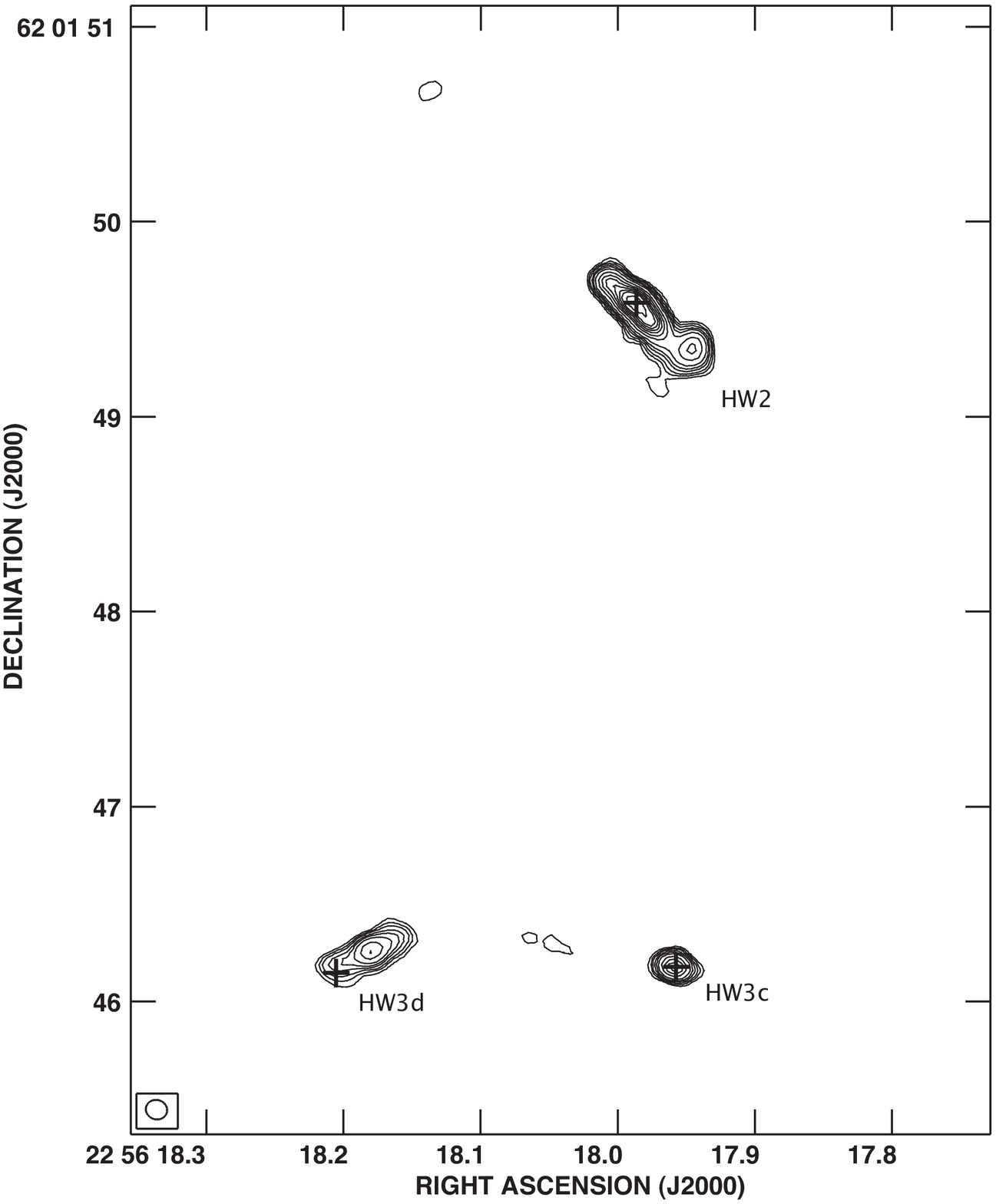}
\end{center}
\caption{Contour map of the continuum emission of Cepheus A made from the February 11, 2006 VLA observations (project number: AC0810). The beam size is 0.11 $\times$ 0.10 arcsecond at the position angle of 73.8$^{\circ}$. Contour levels are $-$5, 5, 7, 9, 12, 15, 20, 30, 40, 50, 60, 70, 80, 100, and 120 times the RMS noise (52 $\mu$Jy beam$^{-1}$) of the map. HW2, HW3d and HWc are shown in the map according to the naming system by Hughes, Cohen, \& Garrington (1995). The plus signs on the HW2, HW3c, and HW3d objects indicate the peak positions of the continuum sources  observed in 1995 (reported by \citealp{tor98}).}

\label{fig: Maser distribution}
\end{figure}

\clearpage


\begin{figure}
\begin{center}
\epsscale{0.8}
\plotone{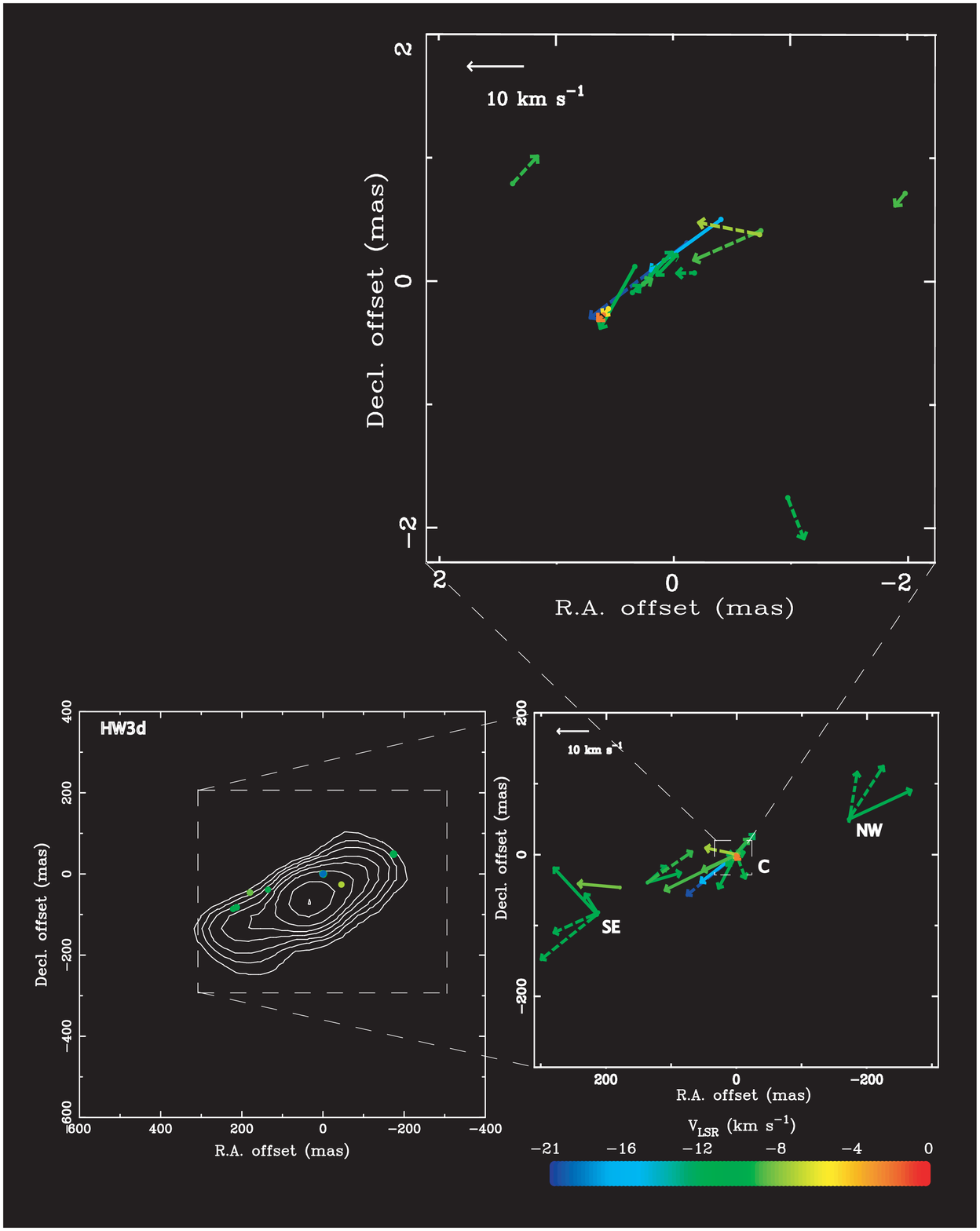}
\end{center}
\caption{Left: The distribution of the H$_{2}$O masers detected in our VERA observations superposed to the HW3d 1.3~cm continuum map obtained with the VLA in 2006. Right: The proper motions of these maser features. The position, length, and direction of an arrow indicate the position, speed, and direction of the maser feature motion on the sky, respectively. The continuous lines represent proper motions traced in 3 or more epochs while the dashed lines represent those traced in 2 epochs only (see also Figure 4 in the Appendix). The motion speed of 10~km~s$^{-1}$ is indicated by a white arrow length in the upper left corner. The color code indicates the LSR velocity of the individual maser features.} 

\label{fig: Proper motions of HW3d H$_{2}$O masers}
\end{figure}


\begin{figure}
\begin{center}
\plotone{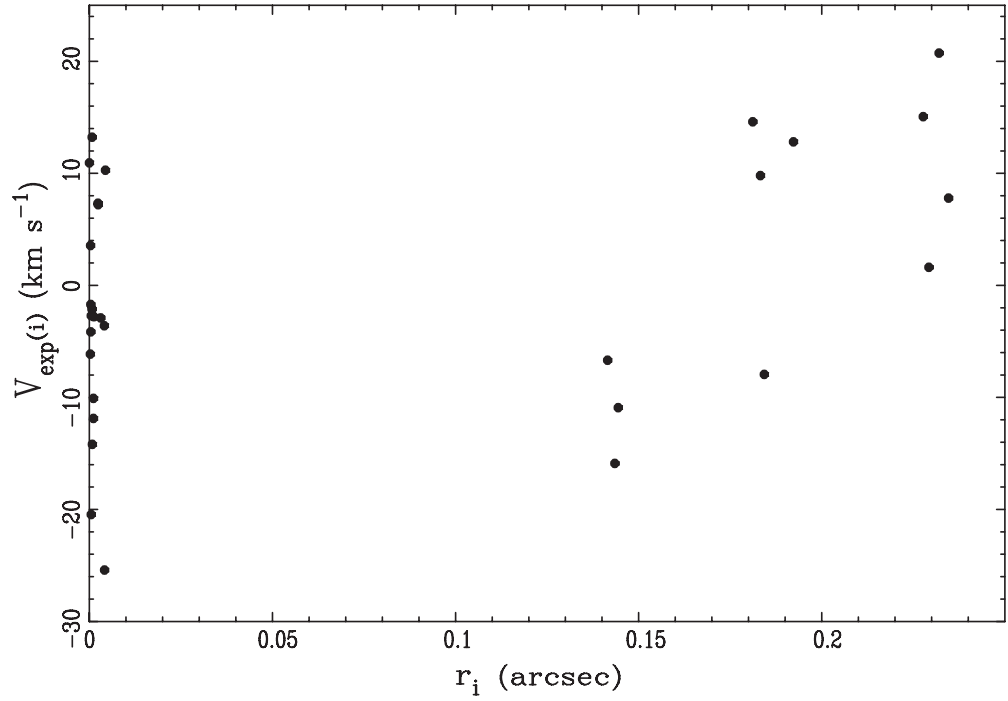}
\end{center}
\caption{Distribution of the expansion velocities of the individual maser features, that were derived from the model fitting in {\it HW3d} (see \S 3.2). The absolute coordinates of the (0,0) position (feature 15, Table 2) are RA(J2000) = $22^h56^m18.1753^s$, DEC(J2000) = $+62^{\circ}01\arcmin46.2114\arcsec$.}
\label{fig: expansion velocities}
\end{figure}







\clearpage

\begin{deluxetable}{lrcrc}
\tablewidth{0pt}
\tablecaption{{VERA observations of Cepheus A.}\label{tab:status}}
\tablewidth{0pt}
\tablehead{
\tableline
\colhead{} & \colhead{Observation\phantom{aaa}}  & \colhead{VERA\tablenotemark{a}}      & 
\colhead{Beam\tablenotemark{b}}    &  \colhead{Duration}  \\
\colhead{\phantom{a}Code}         &   \colhead{Epoch}      &  
\colhead{Noise\tablenotemark{c} [Jy beam$^{-1}$ ]\phantom{aaa}}    &  \colhead{[mas,\hspace{2mm}$^{\circ}$]}  & 
\colhead{Hours}
}
\startdata
r06133b   & 2006 May 13  & 0.12 & 1.49$\times$0.87$\;$, $-$38.8 &   8  \\
r06208an   & 2006 July 27 &  0.40 & 1.38$\times$0.91$\;$, $-$2.9 & 8\\
r06291a  & 2006 October 18  & 0.58 & 1.79$\times$0.85$\;$, $-$40.7 & 11            \\
r06310a & 2006  November 6  & 0.03 & 1.79$\times$0.84$\;$, $-$37.9 & 11          \\
r07004a   & 2007 January 4 & 0.06 & 1.57$\times$0.76$\;$, $-$61.6  & 11             \\
r07049a  & 2007 February 18 & 0.09 & 1.57$\times$0.86$\;$, $-$51.9  & 11              \\
r07103a & 2007 April 13  & 0.09 & 1.64$\times$0.96$\;$, $-$51.4  &11\\
r07135a & 2007 May 15  & 0.10 & 1.46$\times$0.94$\;$, $-$66.0  & 11 \\
r07243a & 2007 August 31  & 0.06 & 1.84$\times$0.81$\;$, $-$82.9 &11 \\
\enddata

\tablenotetext{a}{All the VERA telescope stations (Mizusawa, Iriki, Ogasawara, and Ishigakijima) participated in every epoch of the observations.}
\tablenotetext{b}{Synthesized beam size resulting from naturally weighted visibilities; major and minor axis lengths and position angle.}
\tablenotetext{c}{Average of the typical noise per spectral channel.}

\end{deluxetable}
\clearpage


\begin{deluxetable}{rrccccccccccccccc}
\tabletypesize{\scriptsize}
\rotate
\tablewidth{0pt}
\tablecaption{{Parameters of H$_{2}$O masers associated with HW3d and proper motions measured with VERA.}\label{tab:pmotions_VERA}}
\tablewidth{0pt}
\tablehead{
\multicolumn{1}{c}{}&
\multicolumn{2}{c}{Offset (mas)}&
\multicolumn{4}{c}{Proper motion (mas yr$^{-1}$)} &
\multicolumn{1}{c}{Radial motion (km s$^{-1}$)} &
\multicolumn{9}{c}{Detection at 9 epochs\tablenotemark{a}}\\
\multicolumn{1}{c}{}&
\multicolumn{2}{c}{\hrulefill}&
\multicolumn{4}{c}{\hrulefill}&
\multicolumn{1}{c}{\hrulefill}&
\multicolumn{9}{c}{\hrulefill}\\

\colhead{ID\tablenotemark{b}} & \colhead{R.A.} & \colhead{Dec.} & \colhead{$\mu_{x}$} & \colhead{$\sigma \mu_{x}$} & \colhead{$\mu_{y}$}  & \colhead{$\sigma \mu_{y}$} & \colhead{V$_{\rm LSR}$} & \colhead{1} & \colhead{2} & \colhead{3} & \colhead{4} & \colhead{5} & \colhead{6} & \colhead{7} & \colhead{8} & \colhead{9}\\
}

\startdata
 
 1 &$  -173.61$&$    48.72$&$   3.87$&   1.03 &$   5.12$&   0.16 &$ -10.37$&   -- &-- &-- &-- &+ &-- &+ &-- &-- \\
2 &$  -172.83$&$    50.24$&$  -3.23$&   1.84 &$   4.56$&   0.62 &$ -10.42$&   -- &-- &-- &-- &-- &-- &+ &+ &-- \\
3 &$  -172.29$&$    49.61$&$  -1.87$&   0.36 &$   3.95$&   0.09 &$ -10.00$&  -- &-- &-- &-- &-- &-- &+ &+ &+ \\
4  &$    -3.77$&$     1.52$&$   6.94$&   0.34 &$  -3.16$&   0.22 &$  -9.00$&   -- &-- &+ &+ &+ &-- &-- &-- &-- \\
5  &$    -2.10$&$     0.66$&$   0.48$&   0.08 &$  -0.57$&   0.07 &$  -9.00$&   -- &-- &+ &+ &-- &-- &-- &+ &-- \\
6    &$    -1.13$&$    -1.80$&$  -0.78$&   0.10 &$  -1.91$&   0.11 &$  -9.51$&    + & -- &-- &-- &+ &-- &-- &-- &-- \\
7 &$    -0.90$&$     0.33$&$   2.92$&   3.08 &$   0.56$&   0.80 &$  -7.47$&   -- &-- &-- &-- &-- &-- &+ &+ &-- \\
8 &$    -0.58$&$     0.45$&$   3.37$&   0.25 &$  -2.40$&   0.25 &$ -16.15$&  -- &-- &-- &-- &-- &+ &+ &+ &-- \\
9  &$    -0.36$&$     0.02$&$   0.82$&   0.18 &$  -0.02$&   0.18 &$ -11.47$&  -- &-- &-- &+ &-- &-- &+ &-- &-- \\
10 &$    -0.30$&$     0.26$&$   4.64$&   0.66 &$  -3.51$&   0.41 &$ -20.48$&  -- &-- &-- &-- &+ &+ &-- &-- &-- \\
11  &$     -0.27$&$    0.25$&$   -1.59$&   1.37 &$  1.54$&   1.21 &$  -0.67$&   -- &-- &-- &-- &+ &+ &-- &-- &-- \\
12  &$    -0.21$&$     0.16$&$   0.91$&   0.12 &$  -0.92$&   0.12 &$  -9.82$&   -- &-- &-- &-- &+ &+ &-- &+ &-- \\
13  &$    -0.91$&$     0.36$&$   3.22$&   1.46 &$  -1.38$&   0.58 &$  -8.99$&  -- &-- &-- &-- &-- &+ &+ &-- &-- \\
14  &$    -0.11$&$     0.12$&$  -0.72$&   0.15 &$   0.18$&   0.15 &$  -9.41$&   -- &-- &-- &-- &-- &+ &+ &-- &-- \\
15  &$     0.00$&$     0.00$&$  0.00$&   0.00 &$   0.00$&   0.00 &$  -9.93$&   + &-- &+ &+ &-- &+ &-- &+ &-- \\
16 &$     0.06$&$    -0.07$&$  -0.39$&   0.08 &$   0.22$&   0.06 &$  -9.24$&   -- &-- &+ &+ &+ &+ &-- &+ &-- \\
17  &$     0.11$&$    -0.12$&$  -1.67$&   0.29 &$   1.77$&   0.37 &$ -12.58$&  -- &-- &+ &+ &-- &+ &-- &-- &-- \\
18  &$     0.13$&$     0.07$&$   1.66$&   0.40 &$  -2.89$&   0.63 &$  -9.66$&   -- &-- &+ &+ &+ &-- &-- &-- &-- \\
19  &$     0.15$&$    -0.14$&$  -0.35$&   0.10 &$   0.32$&   0.10 &$ -10.84$&   -- &-- &-- &+ &-- &+ &+ &-- &-- \\
20 &$     0.35$&$    -0.27$&$   0.25$&   0.24 &$  -0.33$&   0.19 &$  -4.79$&   -- &-- &+ &-- &-- &+ &+ &-- &-- \\
21  &$     0.39$&$    -0.32$&$   0.26$&   0.39 &$  -0.26$&   0.35 &$  -1.79$&   -- &-- &-- &+ &-- &+ &-- &-- &-- \\
22  &$     1.14$&$     0.74$&$  -1.20$&   0.68 &$   1.29$&   0.57 &$  -9.33$&   -- &-- &-- &-- &-- &+ &-- &+ &-- \\
23 &$   135.90$&$   -39.14$&$  -1.70$&   0.70 &$   1.37$&   0.42 &$ -12.26$&  -- &-- &-- &-- &+ &+ &-- &-- &-- \\
24  &$   135.85$&$   -39.03$&$  -4.15$&   0.98 &$   2.63$&   1.19 &$  -9.22$&  -- &-- &-- &-- &+ &+ &-- &-- &-- \\
25  &$   136.46$&$   -39.36$&$  -3.12$&   0.07 &$   0.84$&   0.09 &$ -10.00$&  -- &-- &-- &+ &+ &+ &+ &-- &-- \\
26  &$   178.60$&$   -46.23$&$   3.91$&   0.23 &$   0.32$&   0.16 &$  -8.39$&   -- &-- &+ &+ &+ &+ &-- &-- &-- \\
27  &$   211.62$&$   -81.55$&$   4.18$&   0.23 &$  -1.74$&   0.42 &$ -10.63$&   -- &-- &-- &+ &-- &+ &-- &-- &-- \\
28  &$   213.08$&$   -81.99$&$   1.15$&   0.27 &$   1.62$&   0.27 &$ -12.15$&  -- &-- &-- &-- &-- &+ &+ &+ &-- \\
29 &$   213.21$&$   -84.78$&$   4.08$&   0.21 &$   4.04$&   0.29 &$ -10.37$&   -- &-- &-- &-- &+ &-- &+ &+ &-- \\
30 &$   214.91$&$   -85.27$&$   5.18$&   0.42 &$  -3.83$&   1.87 &$ -10.21$&  -- &-- &-- &-- &-- &-- &+ &+ &-- \\

\enddata

\tablenotetext{a}{+ represents detection, while -- represents non-detection in the different epochs.}
\tablenotetext{b}{Maser feature ID number for identification purposes. The absolute coordinates of the (0,0) position (feature 15, Table 2), extrapolated from our astrometric analysis, are RA(J2000) = $22^h56^m18.1753^s$, DEC(J2000) = $+62^{\circ}01\arcmin46.2114\arcsec$.}

\end{deluxetable}
\clearpage


\begin{deluxetable}{ccccccc}
\rotate
\tablewidth{0pt}
\tablecaption{{Position and velocity variance/covariance matrix analyses of the HW3d bipolar outflow.}
\label{tab:pvvcm_VERA}}
\tablewidth{0pt}
\tablehead{
\multicolumn{7}{c}{Diagonalization of the position variance/covariance matrices}\\
\tableline
\colhead{$\psi_{\rm max}$} & & \colhead{$\psi_{\rm min}$} & & \colhead{PA$_{\rm max}$\tablenotemark{a}}\\
\colhead{[mas$^{2}$]} & & \colhead{[mas$^{2}$]} & & \colhead{[$^{\circ}$]}\\
}

\startdata

12697.28  &  & 32.64 & & $108.5$\\

\cutinhead{Diagonalization of the velocity variance/covariance matrices}
\colhead{$\psi_{\rm max}$} & \colhead{$\psi_{\rm mid}$} & \colhead{$\psi_{\rm min}$} & \colhead{PA$_{\rm max}$} & \colhead{PA$_{\rm mid}$\tablenotemark{b}} & \colhead{$\phi_{\rm max}$\tablenotemark{c}} & \colhead{$\phi_{\rm mid}$\tablenotemark{c}}\\

\colhead{[km$^{2}$~s$^{-2}$]} & \colhead{[km$^{2}$~s$^{-2}$]} & \colhead{[km$^{2}$~s$^{-2}$]} & \colhead{[$^{\circ}$]} & \colhead{[$^{\circ}$]} & \colhead{[$^{\circ}$]} & \colhead{[$^{\circ}$]}\\
\tableline

111.41 & 34.90 & 11.66 & $123.0\pm14.5$ & $32.9\pm19.5$ & $-5.3\pm2.5$& $0.9\pm0.4$\\

\enddata

\tablenotetext{a}{Position angle of the largest eigenvalue, ({$\psi_{\rm max}$}).}
\tablenotetext{b}{Position angle of the second largest eigenvalue, ({$\psi_{\rm mid}$}).}
\tablenotetext{c}{$\phi_{\rm max}$ \& $\phi_{\rm mid}$ represent the inclination angle of {$\psi_{\rm max}$} and {$\psi_{\rm mid}$} with respect to the sky plane, respectively.}

\end{deluxetable}
\clearpage

\begin{deluxetable}{cc}
\tablewidth{0pt}
\tablecaption{{Parameters of the best fitted expanding-flow model in the region {\it HW3d} using all the 30 maser features.}\label{tab:models}}
\tablewidth{0pt}
\tablehead{
\colhead{} & \colhead{Systemic proper motion} \\
}

\startdata

$V_{{\rm 0}x}$ (km~s$^{-1}$) & $-$2.5$\pm$3.0\\
$V_{{\rm 0}y}$ (km~s$^{-1}$) & 5.4$\pm$2.2 \\

\cutinhead{Position offset of the reference maser spot}
$x_{\mbox{0}}$ (mas) & 20.0$\pm$10.5 \\
$y_{\mbox{0}}$ (mas) &  $-$17.0$\pm$9.0 \\

$\sqrt{S^{2}}$ \tablenotemark{a} & 3.3

\enddata

\tablenotetext{a}{Mean of the root-mean-square residual of the model fitting (errors at 1$\sigma$).}

\end{deluxetable}
\clearpage

\appendix
\section*{Appendix}

We have carried out careful registration and identification of the H$_{2}$O maser relative proper motions in the HW3d especially for the masers in the C cluster (see Figure 2). Maser features coincident in LSR velocity within 1~km~s$^{-1}$ and position offset within 2~mas between epochs were used in tracing the proper motions. A larger fraction of the proper motions were traced in 3 or more epochs, while some were traced in just 2 epochs. The proper motions traced in 2 epochs are either clearly isolated in LSR velocity or position or in both. This is to avoid any case of proper motion misidentification especially in the complex C maser cluster. Before arriving at the 30 H$_{2}$O maser proper motions, we have carefully dropped some maser proper motions with high identification uncertainties. The tracing of the relative proper motions is shown in Figure 4. A gradient is expected for the R.A. and Declination trace of the maser relative motions except for the reference maser feature (feature 15) which is quasi-stationary, but the LSR velocity is observed to remain constant. It can be observed that the LSR velocity was not actually constant in some of the proper motions. Some of the masers indicate LSR velocity drift, but in this paper we do not consider the possible acceleration of the H$_{2}$O masers.

\begin{figure}
\begin{center}
\epsscale{0.8}
\plotone{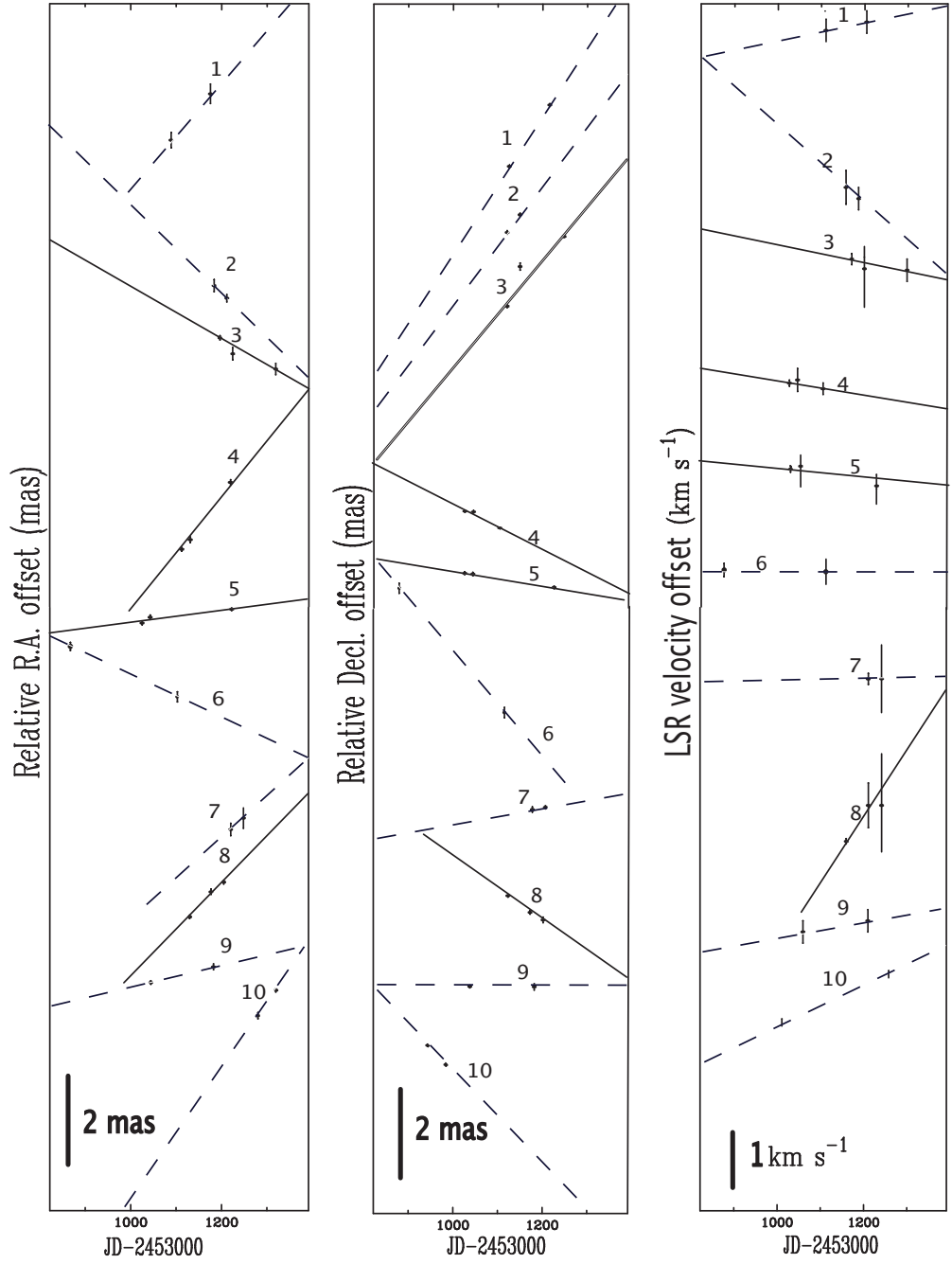}
\end{center}
\caption{continued}
\label{fig: Proper motions}
\end{figure}

\begin{figure}
\begin{center}
\epsscale{0.8}
\plotone{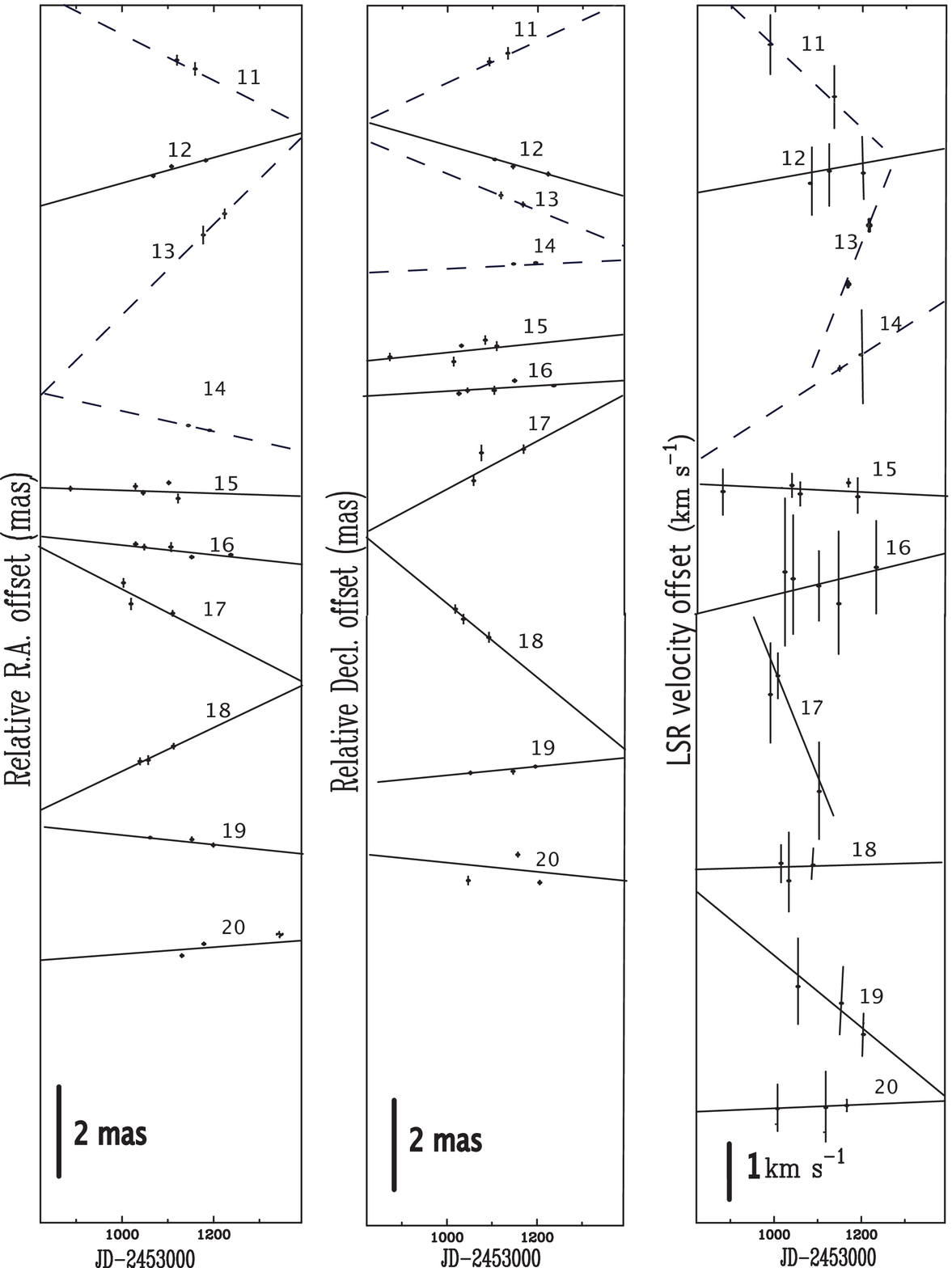}
\end{center}
\addtocounter{figure}{-1}
\caption{continued.}
\end{figure}

\begin{figure}
\begin{center}
\epsscale{0.8}
\plotone{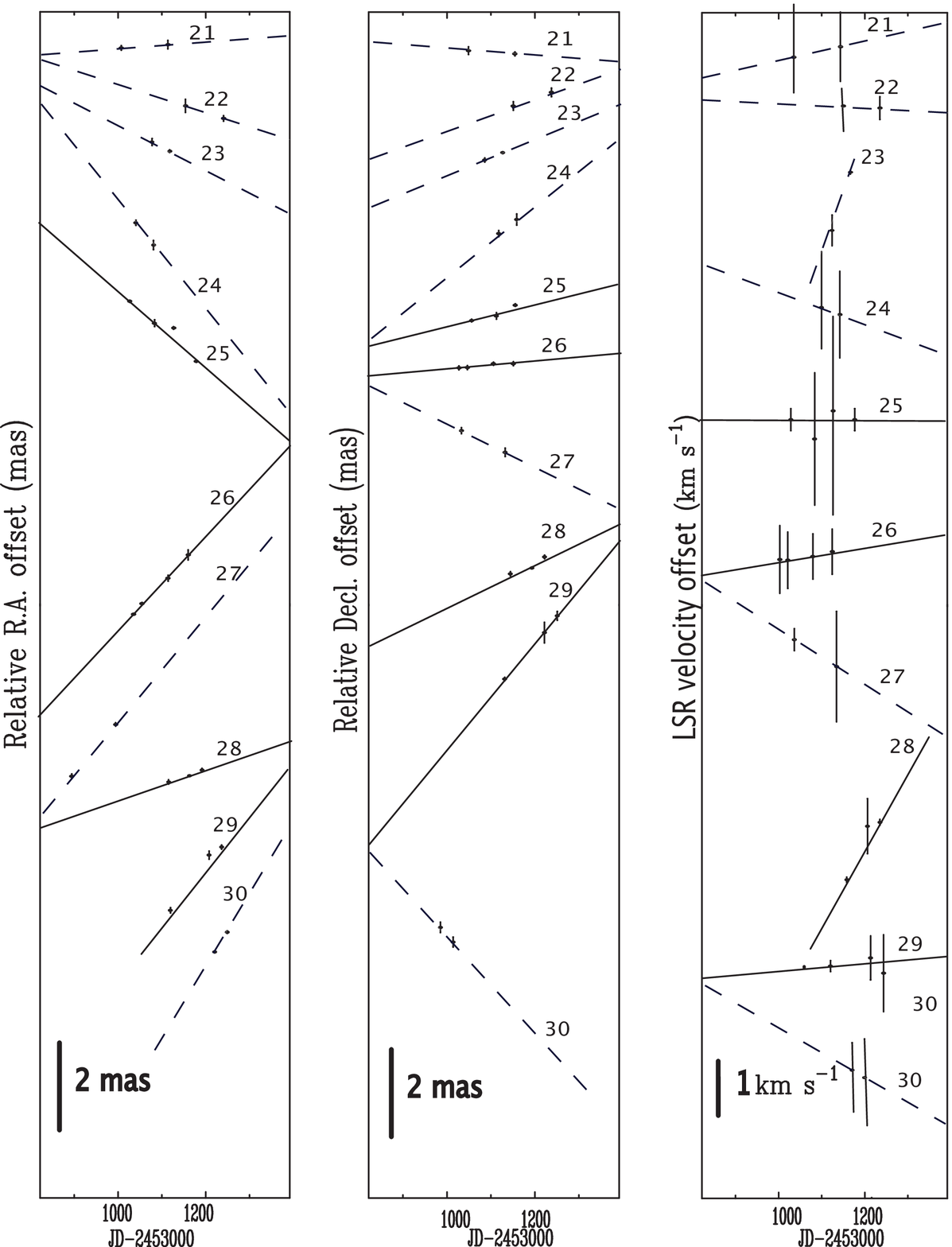}
\end{center}
\addtocounter{figure}{-1}
\caption{Variation of the positions and the LSR velocities of maser features whose relative proper motions were measured. The proper motions were typically obtained in 3 or more epochs (continuous lines), those obtained in only 2 epochs (dashed lines) are either very isolated in velocity and/or position. See the main text for details.}
\end{figure}





\begin{thebibliography}{}


\bibitem[Bloemhof(1993)]{blo93}
Bloemhof, E. E., 1993, ApJ, 406, L75
\bibitem[Bloemhof(2000)]{blo00}
Bloemhof, E. E., 2000, ApJ, 533, 893
\bibitem[Bonnell et al.(1998)]{bon98}
Bonnell, I. A., Bate, M. R., \& Zinnecker, H., 1998, MNRAS, 298, 93
\bibitem[Bonnell \& Bate(2006)]{bon06}
Bonnell, I. A., Bate, M. R., 2006, MNRAS, 370, 488
\bibitem[Cohen et al.(1984)]{coh84}
Cohen, R. J., Rowland, P. R., \& Blair, M. M., 1984, MNRAS, 210, 425
\bibitem[Curiel et al.(2006)]{cur06}
Curiel, S., Ho, P. T. P., Patel, N. A., et al. 2006, ApJ, 638, 878
\bibitem[Dzib et al.(2011)]{dzi11}
Dzib, S., Loinard, L., Rodr\'iguez, L. F., Mioduszewski, A. J., \& Torres, R. M., 2011, ApJ, 733,71
\bibitem[Garay et al.(1996)]{gar96}
Garay, G., Ram\'irez, S., Rodr\'iguez, L. F., Curiel, S., \& Torrelles, J. M., 1996, ApJ, 459,193
\bibitem[Garay et al.(1981)]{gen81}
Genzel, R., Reid, M. J., Moran, J. M., \& Downes, D., 1981, ApJ, 244, 884 
\bibitem[Goddi et al.(2005)]{god05}
Goddi, C., Moscadelli, L., Alef, W., et al. 2005, A\&A, 432, 161
\bibitem[Goddi et al.(2006)]{god06}
Goddi, C., Moscadelli, L., Torrelles, J. M., Uscanga, L., \& Cesaroni, R. 2006, A\&A, 447, L9
\bibitem[Hoare \& Franco(2007)]{hoa07}
Hoare, M. G., \& Franco, J., 2007, in Hartquist T. W., Pittard, J. M., Falle, S. A. E. G., eds, Diffuse Matter from Star Forming Regions to Active Galaxies. Series A\&SSP, 61
\bibitem[Hughes \& Wouterloot(1984)]{hug84}
Hughes, V. A., \& Wouterloot, J. G. A., 1984, ApJ, 276, 204
\bibitem[Hughes \& MacLeod(1993)]{hug93}
Hughes, V. A., \& MacLeod, G. C., 1993, AJ, 105, 1495
\bibitem[Hughes et al.(1995)]{hug95}
Hughes, V. A., Cohen, R. J., Garrington, S., 1995, MNRAS, 272, 469
\bibitem[Imai et al.(2011)]{ima11}
Imai,~H., Omi,~R., Kurayama,~T., et al. 2011, \pasj, 63, 1293 
\bibitem[Imai et al.(2006)]{ima06}
Imai, H., Hirota, T., Umemoto, T., Sorai, K., \& Kondo, T., 2006, PASJ, 58, 883
\bibitem[Imai et al.(2002)]{ima02}
Imai, H., Watanabe, T., Omodaka, T., et al. 2002, PASJ, 54, 741
\bibitem[Imai et al.(2000)]{ima00}
Imai, H., Kameya, O., Sasao, T., et al. 2000, ApJ, 538, 751
\bibitem[Jim\'enez-Serra et al.(2007)]{jim07}
Jim\'enez-Serra, I., Mart\'in-Pintado, J., Rodr\'iguez-Franco, A., et al. 2007, ApJ, 661, L187
\bibitem[Johnson(1957)]{joh57}
Johnson, H. L., 1957, ApJ, 126, 121
\bibitem[Krumholz et al.(2005)]{kru05}
Krumholz, M. R., Klein, R. I., \& McKee, C. F., 2005, in Cesaroni R., Churchwell, E. B., Felli, M. M., Walmsley, C. M., eds, Proc. IAU Symp. 227, Massive Star Birth: A Crossroads of Astrophysics. Cambridge Univ. Press, Cambridge, p. 231
\bibitem[Krumholz et al.(2009)]{kru09}
Krumholz, M. R., Klein R. I., McKee, C. F., Offner, S. S. R., \& Cunningham, A. J., 2009, Science, 323, 754
\bibitem[McKee \& Ostriker(2007)]{mck07}
McKee, C. F., \& Ostriker, E. C., 2007, ARA\&A, 45, 565
\bibitem[McKee \& Tan(2003)]{mck03}
McKee, C., \& Tan, J. C., 2003, ApJ, 585, 850
\bibitem[Moscadelli et al.(2005)]{mos05}
Moscadelli, L., Cesaroni, R., \& Rioja, M. J., 2005, A\&A, 438, 889
\bibitem[Moscadelli et al.(2009)]{mos09}
Moscadelli, L., Reid, M. J., Menten, K. M., et al. 2009, ApJ, 693, 406
\bibitem[Patel et al.(2005)]{pat05}
Patel, N. A., Curiel, S., Sridharan, et al. 2005, Nature, 437, 109
\bibitem[Rodr\'iguez et al.(1994)]{rod94}
Rodr\'iguez, L. F.,  Garay, G., Curiel, S., et al. 1994, ApJ, 430, L65
\bibitem[Surcis et al.(2011)]{sur11}
Surcis, G., Vlemmings, W. H. T., Curiel, S., et al. 2011, A\&A, 527, A48
\bibitem[Torrelles et al.(1998)]{tor98}
Torrelles, J. M., G\'omez, J. F., Garay, G., et al. 1998, ApJ, 509, 262
\bibitem[Torrelles et al.(2001a)]{tor01a}
Torrelles, J. M., Patel, N. A., G\'omez, J. F., et al. 2001a, Nature, 411, 277
\bibitem[Torrelles et al.(2001)]{tor01b}
Torrelles, J. M., Patel, N. A., G\'omez, J. F., et al. 2001b, ApJ, 560, 853
\bibitem[Torrelles et al.(2003)]{tor03}
Torrelles, J. M., Patel, N. A., Anglada, G., et al. 2003, ApJ, 598, L115
\bibitem[Torrelles et al.(2007)]{tor07}
Torrelles, J. M., Patel, N. A., Curiel, S., et al. 2007, ApJ, 666, L37
\bibitem[Torrelles et al.(2011)]{tor11}
Torrelles, J. M., Patel, N. A., Curiel, S., et al. 2011, MNRAS, 410, 627
\bibitem[Vlemmings et al.(2006)]{vle06}
Vlemmings, W. H., T., Diamond, P. J., van Langevelde, H., J., \& Torrelles, J. M., 2006, A\&A, 448,597
\bibitem[Vlemmings et al.(2010)]{vle10}
Vlemmings, W. H., T., Surcis, G., Torstensson, K. J. E., \& van Langevelde, H., J., 2010, MNRAS, 404, 134
\bibitem[Yorke \& Sonnhalter(2002)]{yor02}
Yorke, H. W., \& Sonnhalter, C., 2002, ApJ, 569, 846
\bibitem[Zinnecker \& Yorke(2007)]{zin07}
Zinnecker, H., \& Yorke, H. W., 2007, ARA\&A, 45, 881

\end{thebibliography}
\end{document}